\documentclass[12pt]{article}

\usepackage{amsmath}
\usepackage{amssymb}
\usepackage{changebar}
\usepackage{rotate}
\usepackage{graphics}
\usepackage[dvips]{epsfig}
\usepackage{epsfig}
\usepackage{color} 
\usepackage{graphicx}
\usepackage{pstcol}
\usepackage[sort&compress,square,comma,numbers]{natbib}
\usepackage{seceqn}

\newcommand{\rf}[1]{(\ref{#1})}

\newcommand{\D}{\displaystyle}

\newcommand{\bt}{{\bf t}}
\newcommand{\bn}{{\bf n}}

\newcommand{\eps}{\varepsilon}

\newcommand{\laplace}{\hspace{0em}\bigtriangleup\hspace{-0.1em}}

\newcommand{\pd}[2]{\frac{\partial #1}{\partial #2}}

\newcommand{\pr}{\frac{\partial}{\partial r}}

\newcommand{\pth}{\frac{\partial}{\partial\theta}}

\newcommand{\be}{\begin{equation}}
\newcommand{\ee}{\end{equation}}
\newcommand{\bea}{\begin{eqnarray}}
\newcommand{\eea}{\end{eqnarray}}
\newcommand{\sbea}{\begin{subequations}\begin{eqnarray}}
\newcommand{\seea}{\end{eqnarray}\end{subequations}}
\newcommand{\beas}{\begin{eqnarray*}}
\newcommand{\eeas}{\end{eqnarray*}}
\newcommand{\nn}{\nonumber}

\newcommand{\mywhere}{\quad\mbox{where}\quad}

\begin{document}

\vspace*{1cm}

\section*{Thin film dynamics on a vertically rotating disk 
partially immersed in a liquid bath}

\vspace*{1cm}

\centerline{K. Afanasiev$\,^1$, A. M\"unch$\,^2$, B. Wagner$\,^1$}

\vspace*{1cm}

{\footnotesize{
          $^1$ Weierstrass Institute for Applied Analysis and Stochastics, 
	  \\\,\,\,Mohrenstrasse 39, 10117 Berlin, Germany\\ 
	  $^2$ Institute of Mathematics, Humboldt University, 
	  10099 Berlin, Germany}}
	  
\vspace*{2cm}
	  
\subsection*{\centerline{Abstract}}

\vspace*{1cm}

The axisymmetric flow of a thin liquid film is considered for the
problem of a vertically rotating disk that is partially immersed in
a liquid bath.  A model for the fully three-dimensional free-boundary
problem of the rotating disk, that drags a thin film out of the bath is
set up. From this, a dimension-reduced extended lubrication approximation
that includes the meniscus region is derived.  This problem constitutes a
generalization of the classic drag-out and drag-in problem to the case
of axisymmetric flow.  The resulting nonlinear fourth-order partial
differential equation for the film profile is solved numerically using
a finite element scheme.  For a range of parameters steady states are
found and compared to asymptotic solutions. Patterns of the film profile,
as a function of immersion depth and angular velocity are discussed.

\newpage


\section{Introduction}

The problem of rotating thin film flows has been investigated extensively,
both theoretically and experimentally, due to the many technological
applications.  Starting with the work by Emslie et al. \cite{Emslie58},
various aspects of this type of surface tension driven flow, influencing
the shape and stability of the film, have been studied. These include 
for example non-Newtonian effects the paper by \cite{Fraysse93}, for 
evaporation
\cite{ReisfeldBankoffDavis91} and Coriolis force \cite{MyersCharpin2001}, 
\cite{MyersLombe2006}, see also \cite{Ozaretal2003} for a recent experimental 
study. 
As with those studies most of them dealt with a configuration where the
fluid layer is moving on a horizontally rotating disk. Making use of
the large scale separation between the small thickness of the film and
the length scale of the evolving patterns, thin film models where used
to derive dimension-reduced models of the underlying three-dimensional 
free boundary problems.

For the situation of a disk that is partially immersed in a bath of
liquid and rotating about the horizontal axis, thereby dragging out a
thin film onto the disk (see figure \ref{config}), there are far fewer studies even though this
configuration is typical for many applications.
For example, for oil disk skimmers, which is used as an effective 
device for oil recovery and as an alternative to toxic 
chemical dispersants, used after an offshore oil spill.
Another application is the fluid dynamical aspects in
connection with the synthesis of Poly\-ethy\-len\-ter\-ephtha\-lat(PET)
in polycondensation reactors. These typically consist of a horizontal
cylinder that is partially filled with polymer melt and contains disks
rotating about the horizontal axis of the cylinder, thus picking up and
spreading the melt in form of a thin film over a large area of the disks.

These type of problems always involve a meniscus region, that connects the
thin film to the liquid bath and a spacially oscillating region shortly
before the film is dragged into the bath again. In these two regions
the scale separation is not large anymore and hence the lubrication
approximation is not valid there.  Nevertheless, the meniscus does
play the crucial role of fixing the height of the dragged out film and
therefore both, the meniscus region and similarly the drag-in region
that connecting the thin film to the liquid bath must be accounted for
in a dimension-reduced model, which we will derive in this paper.
The classic and far simpler setting of the free boundary problems for 
falling and rising thin film flows on vertical as well as inclined planes
has been investigated as early as in the work by Landau and
Levich \cite{LL42}.  Their work lead to the prediction of the height
and shape of the thin film emerging out of the meniscus. 
The results were improved by Wilson \cite{Wilson82} and for
the case of a Marangoni-driven rising film by  \cite{schwartz01} 
and M\"unch \cite{c10-muench02}, using
systematic asymptotic analysis in the limit of small capillary numbers. 
Such asymptotic analyses can be applied to more complex 
situations such as the problem we consider here for the vertically rotating disk. 

Previous studies for this problem was performed by 
Christodoulo et al. \cite{CTW90}. The employed the analysis of 
the meniscus region by Wilson 
\cite{Wilson82} for the problem of flow control 
for rotating oil disk skimmers. 
Their study did 
not extend further into  
the thin film region on the remainder of the disk. 
However, for many applications it is important to answer questions 
for example on the maximum surface area of the film 
profile or the optimum volume of liquid that is dragged out and spread on the 
disk, for which the predictions resulting from the simple drag-out study 
will not be sufficient.
Up to
now no complete model for the vertically rotating disk, including its
numerical solution has appeared. This will be the topic of this paper.

In section 2 we set up the corresponding three-dimensional free
boundary problem.  A fully three-dimensional analysis of such flows
represents a very time consuming task, analytically and numerically.
To be able to perform systematic parameter studies we therefore exploit
the large separation of scales to obtain a dimension-reduced lubrication
model. This model will then be extended to match to the flow field in
the meniscus region. For the resulting model we develop in section 3
a weak formulation and a corresponding finite element discretization
for the full dynamical problem. 
Since we want to address here issues of maximum surface area of the film 
profile or the optimum volume of liquid that is dragged out, we are intersted 
mainly in the long-time behavior. This will be the focus of the following 
sections.
In section 4 we solve and discuss for a range of
parameters the emerging steady state solutions. Their shapes near the
meniscus region are then compared to the asymptotic solution of the
corresponding drag-out problem. Away from the liquid bath the film 
profile is compared to asymptotic solutions, by using the methods 
of characteristics. For both regions excellent agreement is found.
Finally, we discuss the novel patterns for the film profile 
as the immersion depth, or the angular velocity is varied. 

\section{Formulation}

\subsection{Governing Equations}
\begin{figure}[h!]
\begin{center}
\unitlength1cm
\begin{picture}(14,8)
\put(7,4.0){\makebox(0,0){
\psfig{figure=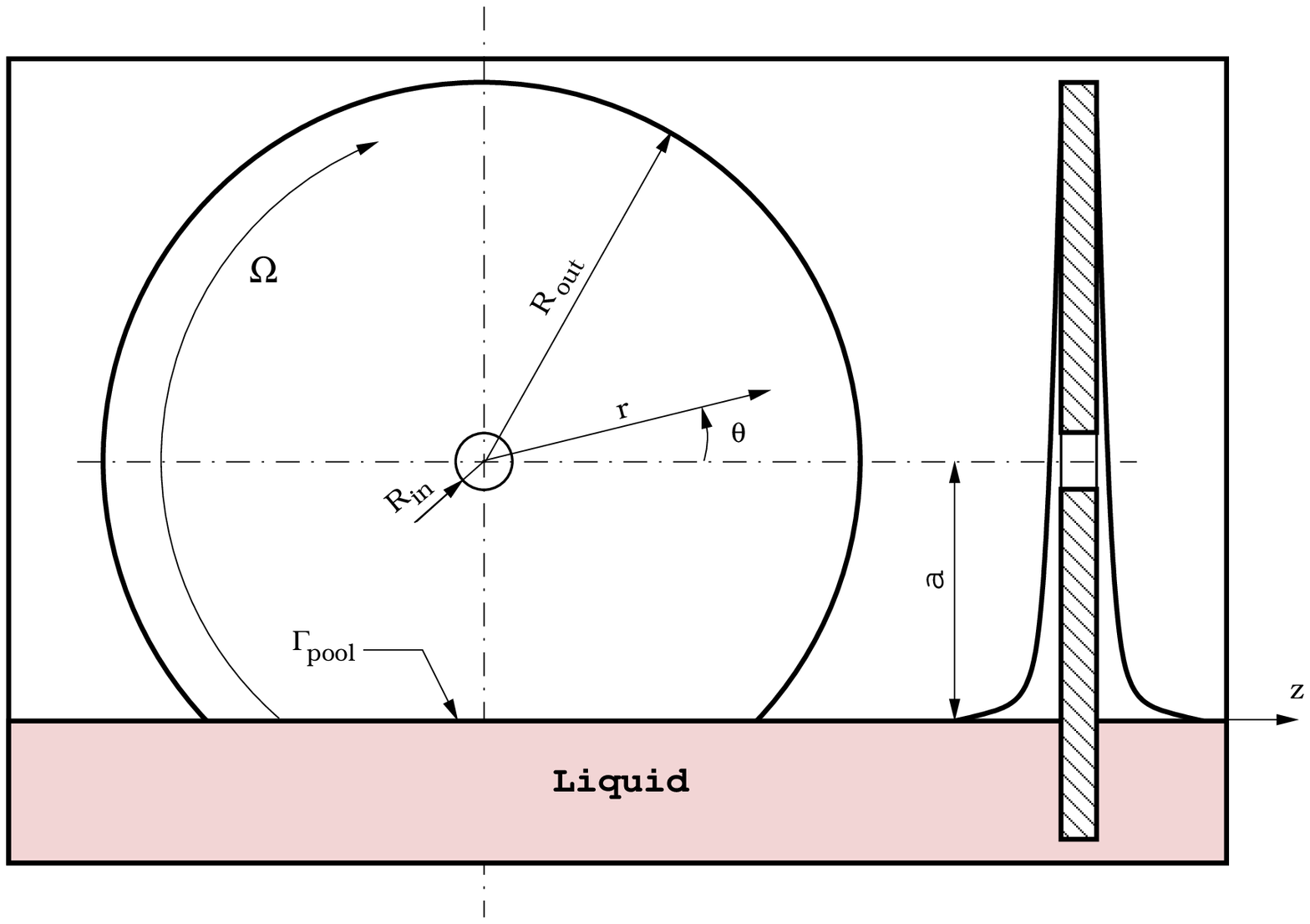,width=11cm,angle=0}}}
\end{picture}
\caption{Configuration of a single disk within a PET-reactor}
\label{config}
\end{center}
\end{figure}

We consider the isothermal flow of an
incompressible, viscous liquid on a vertical disk rotating
in the vertical plane and partially immersed in the liquid. 
We assume that the  disk of radius $R$ rotates with the angular 
velocity $\Omega$ about a horizontal axis, which has distance 
$a$ to the bath, see figure \ref{config}. 

To formulate the problem, we introduce cylindrical polar
coordinates $(r,\theta,z)$ in the laboratory frame of reference.
We let the liquid velocity vector have components $(u,v,w)$ and let $\mathbf{\omega}$ denote the angular velocity vector
with components $(0,0,\Omega)$.  The momentum balance equations can be
expressed as
\sbea 
\hspace*{-0.5cm}\rho\left[u_t+u u_r + \frac{v}{r} u_{\theta} - \frac{v^2}{r} + w u_z\right] 
&=&
-p_r 
+ \mu \left[\laplace u -\frac{2 v_{\theta}}{r^2} -\frac{u}{r^2}
\right]
-\rho g \sin\theta \qquad\nn\\ \label{mom1-dim}\\
\hspace*{-0.5cm}\rho\left[v_t+u v_r + \frac{v}{r} v_{\theta} + \frac{u v}{r} + w v_{z}\right] 
&=&  
-\frac{p_{\theta}}{r } 
+ \mu \left[\laplace v +\frac{2 u_{\theta}}{r^2} -\frac{v}{r^2}
\right]
-\rho g \cos\theta \qquad\nn\\ \label{mom2-dim}\\
\hspace*{-0.5cm}\rho\left[w_t+u w_r + \frac{v}{r} w_{\theta} + w w_z\right] 
&=&
-p_z + \mu \laplace w \qquad\label{mom3-dim}
\seea
where  
\be
\laplace f = \frac{1}{r}\left(rf_r\right)_r +\frac{f_{\theta\theta}}{r^2}+f_{zz}.\label{polar-laplace}
\ee
We let $\rho$, $\mu$ and $p$  denote the density, dynamic 
shear viscosity and the pressure of the liquid, respectively. 
The external force here
is gravity and $g$ denotes the gravitational constant.

The continuity equation is 
\be
\frac{1}{r}(r u)_r +\frac{1}{r} v_{\theta} + w_z=0.\label{cont-dim}
\ee

For the boundary condition at the surface of the disk, i.e. 
$z=0$, that rotates with the velocity $\Omega$,  we 
impose the no-slip condition for $u$ and $v$ and the impermeability 
condition for $w$. Hence, we have 
\be
u=0,\qquad v=r\Omega,\qquad w=0,  \label{bc0-dim}
\ee
respectively.

At the free boundary $z=h(r,\theta,t)$ we require the 
normal stress condition 
\be
\bn\, \Pi \, \bn = 2\sigma\kappa,\label{norm-stress1-dim}
\ee
the tangential stress conditions  
\be
\bn\, \Pi \,\bt_i = 0, \quad\mywhere i=1,2, \label{tang-stress1-dim}
\ee
and the kinematic condition 
\be
h_t = w - u_{|h} h_r - \frac{1}{r} v_{|h}h_{\theta}, \label{kin1-dim}
\ee
which can also be written, upon using the continuity equation, as 
\be
h_t= -\frac{1}{r}\pr\, r \int^h_0 u\,dz - \frac{1}{r}\pth \int^h_0 v\,dz.
\label{kin2-dim}
\ee

The normal and the tangential vectors in radial and angular direction are given by 
\be
\bn=\frac{\left(-h_r, -h_{\theta}/r,1\right)}{\left(1+h_r^2+h_{\theta}^2/r^2\right)^{1/2}},\quad
\bt_1=\frac{\left(1, 0, h_r\right)}{\left(1+h_{\theta}^2/r^2\right)^{1/2}},\quad
\bt_2=\frac{\left(0, 1, h_{\theta}/r\right)}{\left(1+h_{\theta}^2/r^2\right)^{1/2}}, 
\ee
respectively. The stress tensor $\Pi$ is symmetric and has the components 
\bea
\begin{array} {lll}
\Pi_{rr}= -p + 2\mu\, u_r,   
&\Pi_{\theta\theta}= -p + 2\mu\left(\D\frac{v_{\theta}}{r}+\frac{u}{r}\right),
&\Pi_{zz}= -p + 2\mu w_z,\\
\\
\Pi_{r\theta }=\mu\left(\D\frac{u_{\theta}}{r}+v_r-\D\frac{v}{r}\right),
&\Pi_{\theta z}=\mu\left(v_z + \D\frac{w_{\theta}}{r} \right),
&\Pi_{r z}=\mu\left(w_r + u_z\right).
\end{array}
\eea

Finally, we assume surface tension to be constant and denote it by $\sigma$ and the 
mean curvature is given by 
\be
\kappa= \frac12\left(\frac1r\pr\frac{rh_r}{\left(1+h_r^2+h_{\theta}^2/r^2\right)^{1/2}}
+\frac{1}{r}\pth\frac{h_{\theta}/r}{\left(1+h_r^2+h_{\theta}^2/r^2\right)^{1/2}}\right).
\ee

Using this in equations 
\rf{norm-stress1-dim} and \rf{tang-stress1-dim} we obtain the 
boundary conditions for the normal stress
\bea
&&\hspace*{-2cm}
-p+ \frac{2\mu}{1+h^2_r+h_{\theta}^2/r^2}\left[
\left(\frac{u_{\theta}}{r}+v_r-\frac{v}{r}\right)\frac{h_r h_{\theta}}{r}\right.\nn\\
&&
\left.-(w_r+u_z)h_r-\left(v_z+\frac{w_{\theta}}{r}\right)\frac{h_{\theta}}{r} 
+u_rh_r^2+(v_{\theta}+u)\frac{h_{\theta}^2}{r^3}+w_z\right]\label{norm-stress-dim}\\
&&=
\sigma\left[\frac1r\pr\frac{rh_r}{\left(1+h_r^2+h_{\theta}^2/r^2\right)^{1/2}}
+\frac{1}{r}\pth\frac{h_{\theta}/r}{\left(1+h_r^2+h_{\theta}^2/r^2\right)^{1/2}}\right], \nn
\eea
the tangential stress condition in radial direction
\bea
2(w_z-u_r)h_r-\left(\frac{u_{\theta}}{r}+v_r-\frac{v}{r}\right)\frac{h_{\theta}}{r}
&&\label{tang1-stress-dim}\\
&&\hspace*{-3cm}
+(w_r+u_z)(1-h^2_r)-\left(v_z+\frac{w_{\theta}}{r}\right)\frac{h_r h_{\theta}}{r}=0
\nn
\eea
and the tangential stress condition in angular direction 
\bea
2\left(w_z -\frac{v_{\theta}}{r} - \frac{u}{r} \right)\frac{h_{\theta}}{r}
-\left(\frac{u_{\theta}}{r}+v_r-\frac{v}{r}\right)h_r
&&\label{tang2-stress-dim}\\
&&\hspace*{-4cm}
+\left(v_z+\frac{w_{\theta}}{r}\right)\left(1-\frac{h^2_{\theta}}{r^2}\right)-\left(w_r+u_z\right)\frac{h_r h_{\theta}}{r}=0\nn
\eea

\subsection{Lubrication approximation}

The solution of the above 
three-dimensional free boundary problem represents, 
analytically and numerically a time 
consuming task for making accurate parameter studies. 
The key idea that we make use of here in order 
to obtain a mathematically and numerically tractable problem, 
is the exploitation of the scale separation in most parts of this 
flow problem.  

We begin by introducing dimensionsless variables and set 
\begin{equation}
\label{NSGscals}
  \begin{array} {lll}
 r=L\bar{r} ,  & \theta=\bar{\theta},  & z=H\bar{z} , \\
 u=U\bar{u} ,     & v=U\bar{v} ,            & w=W\bar{w}, \\
 p=P\bar{p} , & t=T\bar{t}.  &
  \end{array}
\end{equation}
The characteristic velocity $U$ is set by the velocity of the rotating disk. 
For given radius $R$ of the disk we let 
\be
U=R\Omega.
\ee
We determine the scale for 
the characteristic height $H$ by balancing the dominant viscous term with 
gravitational term in the $u$-momentum equation, which yields 
\be
H=\sqrt{\frac{\mu U}{\rho g}}.
\ee
Furthermore, we require that the pressure must also balance the dominant 
viscous term, so that 
\be
P=\frac{\mu U L}{H^2} 
\ee
and that surface tension is important, so that from the 
normal stress boundary condition we find 
\be
P=\frac{\sigma H}{L^2}.
\ee
This yields the scale for $L$ as 
\be
L=\frac{H}{\D\left(\frac{\mu U}{\sigma}\right)^{1/3}}
\ee
and the time scale is fixed by $T=L/U$.

We assume that the liquid film is very thin and that the velocity in the direction 
normal to the disk is much smaller than along the disk. We let 
\be
\eps=\frac{H}{L} \ll 1
\ee 
be a small parameter and $W=\eps U$. Note that this also means that the capillary 
number Ca is small, 
\be
\mbox{Ca}^{1/3}=\left(\frac{\mu U}{\sigma}\right)^{1/3}=\frac{H}{L}\ll 1 .
\ee

With these scales the non-dimensional equations are 
\sbea 
\hspace*{-0.5cm}\eps^2\mbox{Re}\left[u_t+u u_r + \frac{v}{r} u_{\theta} 
- \frac{v^2}{r} + w u_z\right] 
&=&
-p_r + u_{zz}-\sin\theta \qquad\label{mom1}\\
&&\hspace*{-0.0cm}+ \eps^2 \left[\frac{(r u_r)_r}{r}+\frac{u_{\theta\theta}}{r^2} -\frac{2 v_{\theta}}{r^2} -\frac{u}{r^2}\right]\nn\\
\hspace*{-0.5cm}\eps^2\mbox{Re}\left[v_t+u v_r + \frac{v}{r} v_{\theta} + \frac{u v}{r}+w v_{z}\right] 
&=&  
-\frac{p_{\theta}}{r } + v_{zz} - \cos\theta \qquad\label{mom2}\\
&&\hspace*{-0.0cm}+ \eps^2\left[\frac{(r v_r)_r}{r}+\frac{v_{\theta\theta}}{r^2} +\frac{2 u_{\theta}}{r^2} -\frac{v}{r^2}
\right]\nn\\
\hspace*{-0.5cm}\eps^4\mbox{Re}\left[w_t+u w_r + \frac{v}{r} w_{\theta} + w w_z\right] 
&=&
-p_z + \eps^2 w_{zz} +\eps^4\left[\frac{(r w_r)_r}{r}+\frac{w_{\theta\theta}}{r^2}\right] \qquad\nn\\ \label{mom3}
\seea
where the Reynolds number is  
Re$=\rho\, U L/\mu$ 
and we have dropped the '$\,\,\bar\,$\,'s. 

The boundary conditions at the disk, $z=0$ are 
\be
u=0,\qquad v=\alpha r,\qquad w=0,  \label{bc0}
\ee
where $\alpha=L/R$. 

The boundary conditions 
at the free liquid surface $z=h(r,\theta,t)$ are the conditions 
for normal and tangential stresses 
\bea
&&\hspace*{-2cm}
-p+ \frac{2\eps^2}{1+\eps^2 h^2_r+\eps^2 h_{\theta}^2/r^2}\left[
\eps^2 \left(\frac{u_{\theta}}{r}+v_r-\frac{v}{r}\right)\frac{h_r h_{\theta}}{r}\right.\label{norm-stress}\\
&&
\left.-(\eps^2 w_r+u_z)h_r-\left(v_z+\eps^2\frac{w_{\theta}}{r}\right)\frac{h_{\theta}}{r} 
+\eps^2 u_rh_r^2+\eps^2(v_{\theta}+u)\frac{h_{\theta}^2}{r^3}+w_z\right]\nn\\
&&=
\left[\frac1r\pr\frac{rh_r}{\left(1+\eps^2 h_r^2+\eps^2 h_{\theta}^2/r^2\right)^{1/2}}
+\frac{1}{r}\pth\frac{h_{\theta}/r}{\left(1+\eps^2 h_r^2+\eps^2 h_{\theta}^2/r^2\right)^{1/2}}\right], \nn
\eea

\bea
2\eps^2(w_z-u_r)h_r-\eps^2\left(\frac{u_{\theta}}{r}+v_r-\frac{v}{r}\right)\frac{h_{\theta}}{r}
&&\label{tang1-stress}\\
&&\hspace*{-3cm}
+(\eps^2 w_r+u_z)(1-\eps^2 h^2_r)-\eps^2\left(v_z+\eps^2\frac{w_{\theta}}{r}\right)\frac{h_r h_{\theta}}{r}=0
\nn
\eea
 
\bea
2\eps^2\left(w_z -\frac{v_{\theta}}{r} - \frac{u}{r} \right)\frac{h_{\theta}}{r}
-\eps^2\left(\frac{u_{\theta}}{r}+v_r-\frac{v}{r}\right)h_r
&&\label{tang2-stress}\\
&&\hspace*{-4cm}
+\left(v_z+\eps^2\frac{w_{\theta}}{r}\right)\left(1-\eps^2\frac{h^2_{\theta}}{r^2}\right)-\eps^2\left(\eps^2 w_r+u_z\right)\frac{h_r h_{\theta}}{r}=0,\nn
\eea
and the kinematic boundary condition

\begin{equation}
\frac{\partial h}{\partial t}=
-\frac{1}{r} \frac{\partial }{\partial r} \left(r \int^h_0 u \, \mathrm{d}z \right) 
-\frac{1}{r} \frac{\partial }{\partial \theta} \left( \int^h_0 v \, \mathrm{d}z \right)\,.\label{kinnondim}
\end{equation}

\subsection{Region near the liquid bath}

The scalings introduced so far are appropriate for the thin film region
away from the liquid bath. This yields a leading order theory 
that retains the terms that are dominant for the film profile on the
disk, where slopes are small. Towards the liquid bath the film profile
becomes (in fact infinitely) steep and a lubrication scaling is no
longer appropriate. Rather, the profile is governed by the balance
of gravity and surface tension forces, in fact, much as in a static meniscus.
Hence, the appropriate length scales for all spatial coordinates is
the capillary length scale $l_\mathrm{cap}=\sqrt{\sigma/(\rho g)}$.

This length scale can be easily expressed in terms of the lubrication
length scales $H$ and $L$ times an appropriate power of $\eps$,
so that the new {\em meniscus} length scales (denoted by tildes) become
\begin{equation}
\tilde H = \eps^{-3/2} H,\qquad
\tilde L = \eps^{-1/2} L.
\end{equation} 
The parallel velocity scale is unchanged and equal to $U=R\Omega$,
while the normal is now $U$ instead of $\eps U$.
The time scale 
\begin{equation}
\tilde T = \frac{\tilde L}{U}=\eps^{-1/2} T,
\end{equation}
is again a result of the kinematic condition.  
The pressure scale is determined by surface tension, and we find
\begin{equation}
\tilde P =  \sqrt{\sigma\rho g} =  \eps^{-1/2} P.
\end{equation}
Hence, all variables can be transformed to meniscus scalings simply by
rescaling with powers of $\eps$, according to 
\begin{alignat}{3}
r & = \eps^{-1/2} \tilde r, & \quad
z & = \eps^{-3/2} \tilde z, & \quad
h & = \eps^{-3/2} \tilde h, \notag\\
u & =             \tilde u, & \quad 
v & =             \tilde v, & \quad
w & = \eps^{-1}   \tilde w,  \\
t & = \eps^{-1/2} \tilde t, & \quad 
p & = \eps^{-1/2} \tilde p. & \quad
&\notag
\end{alignat}

Inserting these scalings into \rf{mom1}--\rf{kinnondim}, yields the rescaled equations:
\sbea 
\hspace*{-0.5cm}\eps^{3}\mbox{Re}\left[u_t+u u_r + \frac{v}{r} u_{\theta} 
- \frac{v^2}{r} + w u_z\right] 
&=&
-p_r + \eps^3 u_{zz} - \sin\theta \qquad\label{men:mom1}\\
&&\hspace*{-0.0cm}+ \eps^3 \left[\frac{(r u_r)_r}{r}
  +\frac{u_{\theta\theta}}{r^2} 
  	-\frac{2 v_{\theta}}{r^2} -\frac{u}{r^2}\right],\nn\\
\hspace*{-0.5cm}\eps^{3}\mbox{Re}\left[v_t+u v_r 
+ \frac{v}{r} v_{\theta} + \frac{u v}{r}+w v_{z}\right] 
&=&  
-\frac{p_{\theta}}{r } + \eps^3 v_{zz} - \cos\theta \qquad\label{men:mom2}\\
&&\hspace*{-0.0cm}+ \eps^3\left[\frac{(r v_r)_r}{r}
	+\frac{v_{\theta\theta}}{r^2} +\frac{2 u_{\theta}}{r^2} 
	-\frac{v}{r^2} \right],
	\nn\\
\hspace*{-0.5cm}\eps^{4}\mbox{Re}
	\left[w_t+u w_r + \frac{v}{r} w_{\theta} + w w_z\right] 
&=&
-p_z + \eps^3 w_{zz} 
\qquad\label{men:mom3}
\\
&&\hspace*{-0.0cm}
+\eps^3\left[\frac{(r w_r)_r}{r}
+\frac{w_{\theta\theta}}{r^2}\right] ,
\nn
\seea

where the Reynolds number is  
Re$=\rho\, U L/\mu=\eps^{1/2}\rho\,U\tilde L/\mu=\eps^{1/2}\tilde {\rm Re}$ 
and where we have dropped the '$\,\,\tilde\,$\,'s. 

The boundary conditions at the disk, $z=0$ are 
\be
u=0,\qquad v=\hat\alpha r,\qquad w=0,  \label{men:bc0}
\ee
where $\hat\alpha=\ell_{cap}/R$. 

The boundary conditions for normal and tangential stresses become at $z=h(r,\theta,t)$:
\bea
&&\hspace*{-2cm}
-p+ \frac{2\eps^3}{1+ h^2_r+ h_{\theta}^2/r^2}\left[
 \left(\frac{u_{\theta}}{r}+v_r-\frac{v}{r}\right)\frac{h_r h_{\theta}}{r}\right.\nn\\
&&
\left.-( w_r+u_z)h_r-\left(v_z+\frac{w_{\theta}}{r}\right)\frac{h_{\theta}}{r} 
+ u_rh_r^2+(v_{\theta}+u)\frac{h_{\theta}^2}{r^3}+w_z\right]\label{men:norm-stress}\\
&&=
\left[\frac1r\pr\frac{rh_r}{\left(1+ h_r^2+ h_{\theta}^2/r^2\right)^{1/2}}
+\frac{1}{r}\pth\frac{h_{\theta}/r}{\left(1+ h_r^2+ h_{\theta}^2/r^2\right)^{1/2}}\right], \nn
\eea

\bea
2(w_z-u_r)h_r-\left(\frac{u_{\theta}}{r}+v_r-\frac{v}{r}\right)\frac{h_{\theta}}{r}
&&\label{men:tang1-stress}\\
&&\hspace*{-3cm}
+( w_r+u_z)(1- h^2_r)-\left(v_z+\frac{w_{\theta}}{r}\right)\frac{h_r h_{\theta}}{r}=0,
\nn
\eea
 
\bea
2\left(w_z -\frac{v_{\theta}}{r} - \frac{u}{r} \right)\frac{h_{\theta}}{r}
-\left(\frac{u_{\theta}}{r}+v_r-\frac{v}{r}\right)h_r
&&\label{men:tang2-stress}\\
&&\hspace*{-4cm}
+\left(v_z+\frac{w_{\theta}}{r}\right)\left(1-\frac{h^2_{\theta}}{r^2}\right)
-\left( w_r+u_z\right)\frac{h_r h_{\theta}}{r}=0.\nn
\eea

We now retain all terms that appear to leading order either in the
lubrication or the meniscus scalings. Note that, in the meniscus scalings,
the velocity field decouples to leading order from the pressure field
that determines the surface profile. Hence the dominant terms that govern
$h$ in these scalings consists of the pressure and gravity terms, and of
surface tension, based on the full nonlinear expression for curvature. All
these terms already appear also in the lubrication scaling, except for the
nonlinear curvature. Hence our approximate model retains essentially the
terms from a leading order lubrication theory and the nonlinear curvature
term, i.e., in the bulk we have,
\begin{equation} 
0=-p_r+\eps^3 u_{zz} - \sin\theta, \qquad
0=-\frac{p_\theta}{r}+\eps^3 v_{zz} - \cos\theta,\qquad
0=-{p_z}.
\end{equation}
Boundary conditions at $z=0$ are given by \rf{men:bc0},
and at $z=h$:
\begin{equation} 
-p = 
\left[\frac1r\pr\frac{rh_r}{\left(1+ h_r^2+ h_{\theta}^2/r^2\right)^{1/2}}
+\frac1r\pth\frac{h_{\theta}/r}{\left(1+ h_r^2+ h_{\theta}^2/r^2\right)^{1/2}}\right],
\,
u_z=0,\,
v_z =0.
\end{equation}
Integrating first $p_z=0$ yields a solution that does not depend on 
$z$, and the parallel components for the velocity can 
easily be found to be 
\begin{equation}
u=\eps^{-3} (p_r+\sin\theta)(z^2/2-hz), \qquad
v=\eps^{-3} (p_\theta/r + \cos\theta) (z^2/2-hz)+ \hat \alpha r.
\end{equation}

We plug this into the mass conservation relation \rf{kin2-dim} 
$$
h_t= -\frac{1}{r}\pr\, r \int^h_0 u\,dz - \frac{1}{r}\pth \int^h_0 v\,dz,
$$
and obtain, after rescaling time according to $t=\epsilon^{-3} t'$ (
dropping the prime):
\begin{equation}\label{profeq}
h_t= \frac{1}{r}\pr  \left[r\frac{h^3}{3} (p_r +\sin\theta) \right] 
+ \frac{1}{r}\pth\left[\frac{h^3}{3} (p_\theta/r + \cos\theta) 
	- \hat\Omega r h\right],
\end{equation}
where we have introduced $\hat \Omega =\mu\Omega/\sqrt{\rho g \sigma}$.

Far away from the disk we expect the liquid to be at rest,
since the liquid flow diminishes. The shape of the liquid
is governed by the hydrostatic balance, so that 
the liquid surface is flat, i.e. its curvature is
zero, and it is orthogonal to the direction
of the gravitational force. Therefore 
we require that the function $h$
that describes the surface tends to infinity 
as the far field level of the reservoir is approached ($r\to -a/\sin\theta$),
and the nonlinear curvature tends to zero. 
Therefore, we have 
\be
h(r,\theta)\to\infty, \quad p(r,\theta,t)\to 0\quad\mbox{as}\quad r\to  -a/\sin\theta. \label{bc1reservoir}
\ee

Note at this point, that for the numerical simulation,
the domain has to be cut off at a slightly higher level $r=-a^*/\sin\theta$.
In the numerics, we enforce zero curvature at the cut-off of the domain, 
i.e., set $p=0$ there.

Furthermore, the computational domain 
is intended to be an approximation of a very large reservoir, which
quickly equilibrates any mass change that occurs if liquid is transferred
onto or from the disk. We capture this by setting $h$ to a fixed value ($h=1$) 
at the cut-off $r=-a^*/\sin\theta$. We checked that the results
of the simulations were converged, i.e. changing the value of $h$ 
at the cut-off did not affect the results significantly. 
Summarizing, we use the following conditions for the cut-off domain:
\be
h= 1, \quad p=0 \quad\mbox{at}\quad r=  -a^*/\sin\theta. \label{bc2reservoir}
\ee
We remark here, that in a converged result,
$a^*$ and $a$ are very close values,
so that in the following presentation of the results, we will not 
distinguish these two quantities, and drop the $*$.



For the boundary conditions towards the inner ($r=R_{in}$) and 
outer ($r=R_{out}$) confinements 
of the disk we assume that both, 
the reservoir and the disk to be enclosed in a cylinder
that tightly surrounds the disk, which is an arrangement that reflects 
the typical situation inside a PET reactor. 
No liquid is injected or lost at the (solid) axis, nor through the
cylinder; therefore, we set the mass flux to zero, which is the significance
of the following boundary conditions 
\bea
p_r+sin \theta=0, \quad \mbox{as}\quad r\to R_{in}, R_{out}. \label{bcnat1}
\eea
A second condition is required. At the axis, the liquid surface meets
the solid axis, forming a contact-line, where it is natural to impose
a contact angle condition. 
We set the contact angle to $90^\circ$, i.e.\ $h_r=0$, which is  
the easiest value to implement in our finite element approach. Hence, we have 
\bea
h_r=0,             \quad \mbox{as}\quad r\to R_{in}, R_{out}.\label{bcnat2}
\eea


\section{Numerical method}

We now give a brief description of the numerical method used to solve problem 
\rf{profeq}-\rf{bcnat2}.
The meniscus equations may be rewritten for simlicity as
\begin{eqnarray} \label{simplh}
r \; \frac{\partial h}{\partial t}&=&   \frac{\partial Q^r }{\partial r}
                                    + \frac{\partial Q^{\theta} }{\partial \theta}, \\
\label{simplp}
-\frac{1}{2} \; r \; p&=&\frac{\partial q^r }{\partial r}
      +\frac{\partial q^{\theta} }{\partial \theta},
\end{eqnarray}
where fluxes $Q^r,q^r$ and $Q^{\theta},q^{\theta}$ in $r$ and 
$\theta$ directions are defined as 
\begin{eqnarray}
Q^r &=&r  \frac{h^3}{3} (p_r+ \sin \theta), \hspace*{1.9cm}
q^r=  \frac{r h_r }{\sqrt{1+ h^2_{r}+h^2_{\theta}/r^2}}, \\
Q^{\theta}&=&\frac{h^3}{3} (\frac{1}{r}p_{\theta} + \cos \theta) + r\Omega h, 
\quad
q^{\theta}= \frac{h_{\theta}}{r \sqrt{1+ h^2_{r}+h^2_{\theta}/r^2}},
\end{eqnarray}
respectively.
For the outlet boundary condition we take natural boundary condition, i.e. 
zero fluxes in the direction of a normal vector.
\begin{align} 
&Q^r(r,\theta,t)=0,            &  r \rightarrow R_{out}, \label{outlet1}\\
&q^r(r,\theta,t)=0,            &  r \rightarrow R_{out}.\label{outlet2}
\end{align}
Similarly, we choose for the conditions towards the origin the 
natural boundary conditions 
\begin{align} 
&Q^r(r,\theta,t)=0,            &  r \rightarrow R_{in}, \label{originME1}\\
&q^r(r,\theta,t)=0,     &  r \rightarrow R_{in}.\label{originME2}
\end{align}
For the immersing boundary condition, where the thin film connects to the 
liquid bath we let the curvature of the free surface  
vanish. Hence, we require the boundary conditions 
\rf{bc1reservoir} and \rf{bc2reservoir}. 

The weak formulation for (\ref{simplh}, \ref{simplp}) under the boundary conditions \rf{outlet1}-\rf{originME2} and (\ref{bc1reservoir}, \ref{bc2reservoir}) can be derived
by multiplying (\ref{simplh}) and (\ref{simplp}) by a suitable test function $\phi$, integrating over the domain  \\
$\Lambda=\left\{ (r,\theta): R_{in}<r<R_{out}\;\, \mbox{and}\; r\sin\theta>-a\right\}$
and evaluating at the boundary.
Then the weak formulation of the boundary value problem for (\ref{simplh},\ref{simplp}) requires us to seek $(h,p) \in H^1(\Lambda)$, such that
\begin{eqnarray}
\int\limits_{\Lambda} r \frac{\partial h}{\partial t} \; \phi      \;   \mbox{d} \Lambda
= 
  - \int\limits_{\Lambda} \left(  Q^r  \frac{\partial \phi} {\partial r}
  +  Q^{\theta}  \frac{\partial \phi }{\partial \theta}   \right)       \;   \mbox{d} \Lambda 
  + \int\limits_{\Gamma} Q^r  \; \phi \; n_r                        \;    \mbox{d} \Gamma                  
  \\
\frac{1}{2} \int\limits_{\Lambda} r \; p \; \phi                                   \;    \mbox{d} \Lambda
= 
\int\limits_{\Lambda} \left(  q^r \;   \frac{\partial \phi }{\partial r}  
                             + q^{\theta} \frac{\partial \phi }{\partial\theta }   \right)  \;    \mbox{d} \Lambda
- \int\limits_{\Gamma}   q^r \phi \; n_r                               \;    \mbox{d} \Gamma
\end{eqnarray}
for all functions $\phi \in V=W^1_2(\Lambda)$. 
Respecting the boundary conditions $ p_r=0, \; h_r=0$, the following integral equations 
\begin{eqnarray}
\label{weakF1}
\int\limits_{\Lambda} r \frac{\partial h}{\partial t} \; \phi      \;   \mbox{d} \Lambda
= 
  - \int\limits_{\Lambda} \left(  Q^r  \frac{\partial \phi} {\partial r}
  +  Q^{\theta}  \frac{\partial \phi }{\partial \theta}   \right)       \;   \mbox{d} \Lambda 
  + \int\limits_{\Gamma} \left(  \frac{r h^3}{3} B cos \theta \; \phi   \right)    \;    \mbox{d} \Gamma        \\
  \label{weakF2}
\frac{1}{2} \int\limits_{\Lambda}r \; p \; \phi                                   \;    \mbox{d} \Lambda
= 
\int\limits_{\Lambda}   \left(  q^r \;   \frac{\partial \phi }{\partial r}  
                             + q^{\theta} \frac{\partial \phi }{\partial\theta }   \right)    \;    \mbox{d} \Lambda
\end{eqnarray}
will now be discretised.
For the discretisation of the problem  we devide the domain $\Lambda$
in non-overlapping triangular elements $\Lambda_e$ and  replace $H^1(\Lambda)$ and $V(\Lambda)$ by finite dimensional
 subspaces $S$ and $V^h$, respectively.
We also choose $\phi=\phi_i,\; i=1,2,\ldots,N$ with $N$ denoting the number of nodes in the element $\Lambda_e$ and let 
\begin{eqnarray}
\label{FEapprox1} 
h_{e}(r,\theta,t)=\sum\limits_{i=1,N} h_i(t) \phi_i(r,\theta) \\
\label{FEapprox2}
p_{e}(r,\theta,t)=\sum\limits_{i=1,N} p_i(t) \phi_i(r,\theta)
\end{eqnarray} 
be the functions that approximate $h$ and $p$ on this element, respectively.
The domain integrals can now be replaced by the sum of integrals taken 
separatly over the elements of triangulation.

The details of the finite element scheme is described in the following section.

\subsection*{Finite element scheme}

\begin{figure}[h!]
\begin{center}
\unitlength0.9cm
\begin{minipage}[b]{5cm}
\begin{eqnarray*}
\phi_1&=& L_1(2L_1-1) \\
\phi_2&=& L_2(2L_2-1) \\
\phi_2&=& L_3(2L_3-1) \\
\phi_4&=& 4 L_1 L_2 \\
\phi_5&=& 4 L_2 L_3 \\
\phi_6&=& 4 L_3 L_1 
\end{eqnarray*} 
\end{minipage} 
\begin{picture}(7,5.0)
\put(4.5,2.2){\makebox(0,0){
\psfig{figure=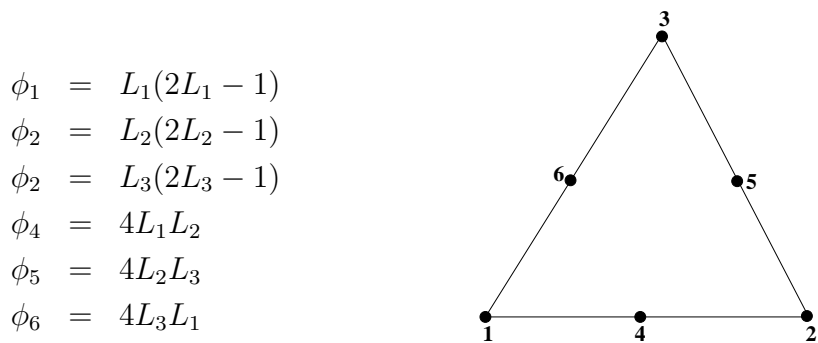,width=4.5cm,angle=0}
}}
\end{picture}
\caption{Basic functions for a six node quadratic triangular element}\label{sixT}
\end{center}
\end{figure}

Let the time interval $[0,T]$ be subdivided into intervals with the time step $\tau$,
$t_n=t_{n-1}+\tau, \; n=1,2,\ldots,N_T $ and denote
$$h^n=\left(
\begin{array}{*{1}{c}}
h_1(t^n)  \\
h_2(t^n)  \\
\vdots    \\
h_N(t^n)  \\
\end{array}
\right),   \qquad
p^n=\left(
\begin{array}{*{1}{c}}
p_1(t^n)  \\
p_2(t^n)  \\
\vdots    \\
p_N(t^n)  \\
\end{array}
\right).
 $$
By substitution of equations (\ref{FEapprox1}, \ref{FEapprox2}) into the weak formulation and its implicit backward Euler discretisation, expressions (\ref{weakF1}, \ref{weakF2}) can be written in matrix notation as the 
following finite nonlinear system
\begin{eqnarray}
L h^{n+1} + \tau \left[ C^r  g^h_1(h^{n+1},p^{n+1}) + C^{\theta} g^h_2(h^{n+1},p^{n+1}) +s(h^{n+1}) \right] =  L h^n , \\
L p^{n+1} = 2 \left[  C^r g^p_1(h^{n+1}) +C^{\theta} g^p_2(h^{n+1}) \right]  
\end{eqnarray} 
where matrices and vectors are defined by 
\begin{eqnarray}
\label{1a} 
L_{ij}&=& \int\limits_{\Lambda_e} r \, \phi_i    \, \phi_j \;\mbox{d}\Lambda                              \\
\label{2a} 
C^r_{ij}&=& \int\limits_{\Lambda_e}  \frac{\partial \phi_i}{\partial r}    \, \phi_j \;\mbox{d}\Lambda  \\
\label{3a} 
C^{\theta}_{ij}&=& \int\limits_{\Lambda_e}  \frac{\partial \phi_i}{\partial  \theta}    \, \phi_j \;\mbox{d}\Lambda  \\
\label{4a} 
M_{ij}&=& \int\limits_{\Lambda_e}  \phi_i    \, \phi_j \;\mbox{d}\Lambda   \\
\label{5a} 
g^h_1 &=& \frac{r w}{3} \left( q^p_r +B \; cos \theta \right) ,                                  \\
\label{6a} 
g^h_2 &=& \frac{w}{3} \left( \frac{q^p_{\theta}}{r} -B \; sin \theta +r \,\Lambda \, h \right)  ,         \\
\label{7a} 
g^p_1 &=& \frac{r q^h_r}{(1+(q^h_r)^2 +(q^h_{\theta})^2/r^2)^{\frac{1}{2}}} ,                 \\
\label{8a} 
g^p_2 &=& \frac{q^h_{\theta}}{r^2(1+(q^h_r)^2 +(q^h_{\theta})^2/r^2)^{\frac{1}{2}}} ,
\end{eqnarray}
\begin{eqnarray}       
\label{9a} 
w     &=&  M^{-1} a, \\
\label{1b} 
a_{i} &=& \sum_{m,l,j} h_m\, h_l\, h_j \,
    \int\limits_{\Lambda}
    \phi_m  \, \phi_l  \, \phi_j   \, \phi_i         \; \mbox{d}\Lambda,               \\
    \label{2b} 
Y^r          &=& M^{-1}(C^r)^T ,     \qquad   Y^{\theta}=M^{-1}(C^{\theta})^T,  \\
\label{3b} 
q^h_r        &=&  Y^r        \, h,   \qquad   q^h_{\theta} =  Y^{\theta} \, h, \\
\label{4b} 
q^p_r        &=&  Y^r        \, p,   \qquad   q^p_{\theta} =  Y^{\theta} \, p, \\
\label{b5}  
s_i          &=&
 \begin{cases}  0, & \quad \Gamma_e := \Lambda_e \bigcap \Gamma = 0, \\
               \int\limits_{\Gamma_e}
       \left(  \frac{h^3_i}{3} B \; r \; cos \theta \right) \; \mbox{d}  \Gamma,
                   &  \quad  \Gamma_e \neq 0.
 \end{cases}
\end{eqnarray} 
\paragraph{Evaluation of matrix and vector coefficients}
The various element matricies and vectors expressed by the equations above are spatial integrals 
of the various interpolation functions and their derivatives.
These integrals can be evaluated analytically. The remaining ones are 
obtained using numerical quadrature procedure.
Matrix and vector coefficients for triangular elements are evaluated 
using a seven-point quadrature scheme for quadratic triangles.
\paragraph{Triangulation}
We use the six node quadratic triangular elements as shown in Figure \ref{sixT} 
and following basic functions written in the so called natural coordinates $L_i, i=1,2,3$  based on area ratios (see in \cite{FemBem}).
The grids were generated by using the automatic mesh generator \cite{Reinelt} based upon the Delaunay refinement algorithm.
\paragraph{Assembling the global equation system}
The contributions of the element coefficient matrices and vectors  \rf{1a}-\rf{b5}  are added by the common global node
for the assembling of the global nonlinear equation system similar to \cite{SchwarzFEM}.
The global equation system can be written in the form
\begin{equation}
R(U)=F,
\end{equation}
where $U$ is constructed from the vectors $h^{n+1}$ and $p^{n+1}$ in all grid nodes.
\paragraph{Time stepping}
The time stepping algorithm is customarily implemented with a Newton-Raphson equilibrium iteration loop.
In the each time step the following nonlinear problem must be solved
\begin{equation}
G(U):=R(U)-F=0.
\end{equation}
The linearized equation can be written on the basis of the Taylor expansion 
$$G(U_{i+1})=G(U_i)+ \underbrace{\frac{\partial G}{\partial U} \bigg|_{U=U_i}}_{K(U_i)} \triangle U_{i+1}. $$
At each step of Newton's method, some direct or iterative method must 
be used to solve the large linear algebra problem produced by the two-dimensional linearized operator 
\begin{eqnarray}
K(U_i)  \, \triangle U_{i+1} = - G(U_i) \\
\mbox{with} \qquad U_{i+1} = U_i + \triangle U_{i+1}
\end{eqnarray}
Here, we find it convenient to use the non-symmetric multi-frontal 
method for large sparse linear systems from the packet UMFPACK \cite{UMFPACK}.

\section{Steady states}
\subsection{The half-immersed disk}

We first discuss the case for the half-immersed disk at some length and 
compare our numerical results to asymptotic solutions near the 
liquid bath and in the thin film region of the disk before we 
consider different immersion depths. 
\subsubsection{Numerical results}
We consider now a disk
rotating about the horizontal axis with a constant angular speed $\Omega$ and
being half-immersed in the liquid bath. 
The triangulation of the computational domain is performed in cylindrical coordinates $r,\theta$.
The finite element mesh, used here, consists of 9533 triangular elements and 19522 nodes.
The mesh is refined on the boundary $\Gamma_{pool}$
to resolve the meniscus region.
The steady state for equations (\ref{simplh}, \ref{simplp}) is obtained via 
time integration with an adaptive time step. 
As the stopping criterion for the Newton iterations a general threshold for the residuum
$||G(U)||< 10^{-13}$ is applied. 

\begin{figure}\centering
\includegraphics*[width=0.45\textwidth,angle=270]{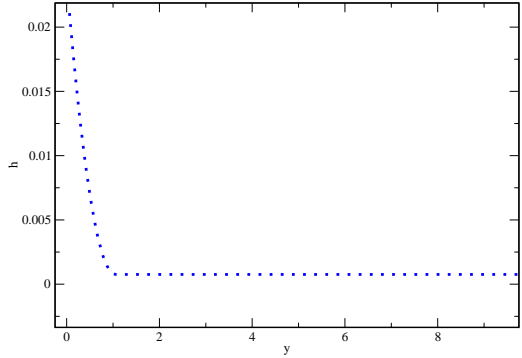}
\caption{Initial state.}
\label{fig:IC} 
\end{figure}

\begin{figure}\centering
\includegraphics*[width=0.5\textwidth, height=0.3\textheight]{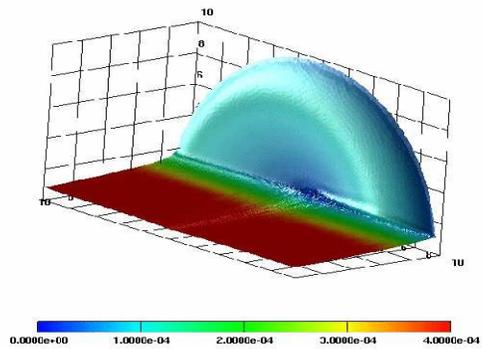}
\includegraphics*[width=0.5\textwidth, height=0.3\textheight]{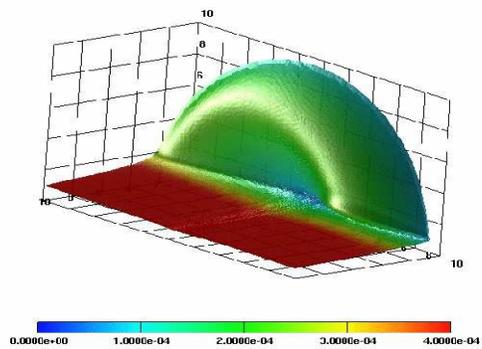}
\includegraphics*[width=0.5\textwidth, height=0.3\textheight]{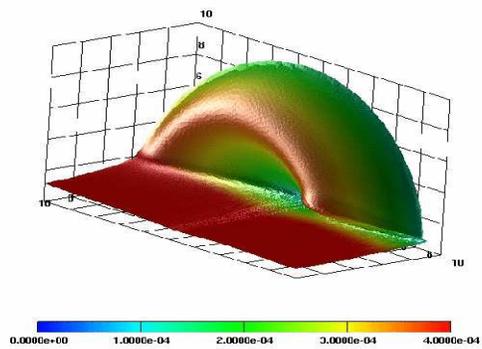}
\caption{Steady solutions at
$\Omega=0.277, 1.0, 2.0 \,$ r.p.m.}
\label{fig:HK1}
\end{figure}

Without loss of generality we choose the following values for the 
parameters throughout the paper: 
$\mu= 1\,\mbox{Pa s}$, $\rho=1000\,\mbox{kg/m$^3$}$, $\sigma=72.7 e-3\,\mbox{N/m}$, $R= 2.723e-2\,\mbox{m}$,  
$g=9.81\,\mbox{m/s$^2$}$.  
For the initial state we choose a partially constant profile on the top of the disk and a partially parabolic
one towards the liquid bath, as shown in the Figure \ref{fig:IC}.
For the given $U=7.917e-4 \,\mbox{m/s}$ and $R$ we find the angular velocity 
$\Omega=U/R=0.02908 \mbox{s}^{-1}= 0.277 \, \mbox{r.p.m}$,
where the last equality is obtained  by multiplying with 
$60/2\pi$.
For these values the capillary number is small $\mbox{Ca}=0.01089$, 
the length scales are 
$\tilde H=\tilde L=l_{\mathrm{cap}}=(\sigma/\rho g)^{1/2}=2.723e-3
\mbox{m}$, $\hat{\Omega}=1.089e-3$ and the dimensionless radius 
of the disk is $10$. The resulting steady state is shown on the top of 
figure \ref{fig:IC}. It shows the an increase of thickness as the 
radius and hence the angular velocity increases, as expected. With 
increasing $\theta$ gravity causes the film to move downwards resulting in 
a ridge of fluid, that thickens with increasing $\theta$ and 
reenters the bath with a typical capillary ridge. 
If we increase $\Omega$, while keeping the other
dimensional parameters and disk radius fixed, neither the
length scales nor the dimensionless radius
of the disk change, but $U$, the capillary number and $\hat{\Omega}$ do. 
As examples we let 
\begin{eqnarray*}
\Omega=1.0\quad\mbox{rpm}: & U=  0.2851e-2\quad\mbox{m/s}, \quad \mbox{Ca}=0.03922, \quad \hat{\Omega}=3.922e-3,\\
\Omega=2.0\quad\mbox{rpm}:  & U=0.5703e-2\quad \mbox{m/s}, \quad \mbox{Ca}=0.07845, \quad \hat{\Omega}=7.843e-3,\\
\end{eqnarray*}
Figure \ref{fig:HK1} 
illustrates the steady states for rotation velocities  
$\Omega=0.277$, $1.0$, $2.0$ r.p.m. from top to bottom, respectively. 
In all 
cases the height of the film in the figures is multiplied by the factor 
of $10$ to contrast more clearly the structure of the 
film patterns. 

\begin{figure}
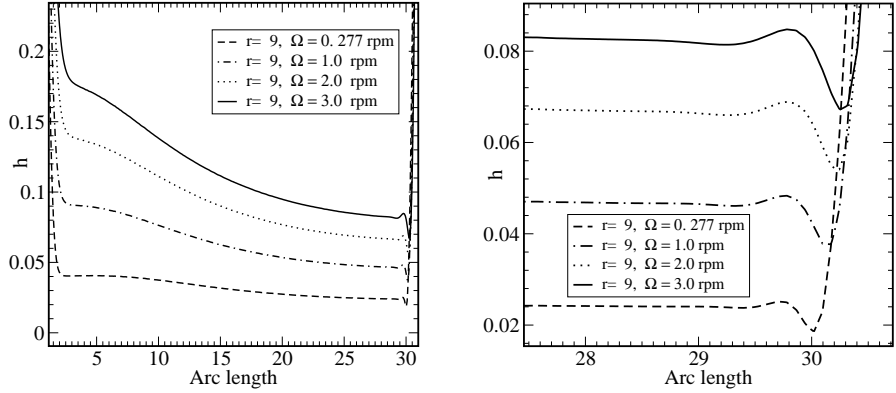
\centering
\includegraphics*[width=0.4\textwidth]{h-fem-r9-sc-many.eps}\qquad
\includegraphics*[width=0.4\textwidth]{h-fem-r9-sc-many-dip.eps}
\caption{Comparisons of the profiles of the cross sections for the film profile 
for radius $r=9$. On the right hand side where the film is pulled into the bath 
a capillary wave is formed. This region is enlarged on the right figure. 
Note, ``Arc length'' denotes $r\theta$.}
\label{fig:fem-cross-r9} 
\end{figure}

One observes for all values of $\Omega$ of the steady solutions 
a region of liquid drag-out with a meniscus profile and a drag-in region with 
a capillary wave on the opposite side of the axis. 
Such an oscillation of the height is typically found for the {\it reverse } 
Landau-Levich problem when a liquid thin film is dragged into a liquid 
bath, see for example \cite{cc73, Kheshgi92, ruschak78, WJ83}.
It can be seen more clearly when comparing the cross sections 
of the liquid profiles at constant radii. 
In figure \ref{fig:fem-cross-r9} we compare 
for the radius $r=9$ the cross section 
for $\Omega=0.277$, $1.0$, $2.0$, $3.0 \,$ r.p.m.
(Note, that here as further below, values such as for $r$ 
without an explicit dimensions are in fact dimensionless). 
The figure also shows that the the average liquid height increases when 
$\Omega$ increases. 

These results are qualitatively in accordance with the 
problem for the drag-out and drag-in cases. We will further 
investigate the quantitative comparison. 

\subsubsection{Comparison with the drag-out problem }

\paragraph{Asymptotic estimate of the film thickness}

We now derive an asymptotic approximation of the film thickness using a one dimensional approximation based on
the results of Landau, Levich \cite{LL42} and Wilson \cite{Wilson82} for the planar-symmetric case.

For our comparisons we focus on the case where the disk is half 
immersed, i.e.\ $a=0$.
Then, if we only retain the axial components in the stationary
form of \rf{profeq}, and after substituting
$r\theta \mapsto y$, $r d\theta \mapsto dy$ we obtain the equation
\begin{equation}\label{h}
\frac{d}{d y} \left[\frac{h^3}{3} (p_y + 1) 
	- \hat\Omega r h\right]=0,
\end{equation}
with
\begin{equation}\label{p}
p=-\frac{d}{dy} \frac{h_y}{(1+h_y^2)^{1/2}}.
\end{equation}

Boundary conditions are
\begin{equation}\label{bc1}
\lim_{y\rightarrow\infty} h = h_\infty, \qquad
\lim_{y\rightarrow 0} h = \infty, \quad \lim_{y\rightarrow 0} p = 0.
\end{equation}
Integrating \rf{h}, \rf{p} once and using the boundary conditions~\rf{bc1}
yields
\begin{equation}\label{stateq}
h^3\frac{d^2}{d y^2}\frac{h_y}{(1+h_y^2)^{1/2}}
= -3r\hat \Omega (h-h_\infty) + (h^3-h_\infty^3).
\end{equation}
We rescale this equation to bring it into the form 

\begin{equation}\label{resc}
h=(r\hat\Omega)^{1/2} \bar h, \quad
h_\infty=(r\hat\Omega)^{1/2} \bar h_\infty, \quad
y =(r\hat\Omega)^{1/6} \bar y,  
\end{equation}
to get
\begin{equation}\label{stateq2}
\bar h^3\frac{d^2}{d \bar y^2}
\frac{\bar h_{\bar y}}{\left(1+(r\hat\Omega)^{2/3}{\bar h}_{\bar y}^2 \right)^{1/2}}
= -3(\bar h-\bar h_\infty) + (\bar h^3-\bar h_\infty^3).
\end{equation}
For this equation, Wilson's formula \cite{Wilson82} gives the asymptotic approximation for the film thickness
$$
\bar h_\infty 
=0.94581 \, (r\hat\Omega)^{1/6},
$$
i.e. from~\rf{resc}, 
\begin{equation}\label{hinfty2}
h_\infty=0.94581 \, (r\hat\Omega)^{2/3}.
\end{equation}

\begin{figure}\centering
\includegraphics*[width=0.7\textwidth, height=0.55\textwidth]{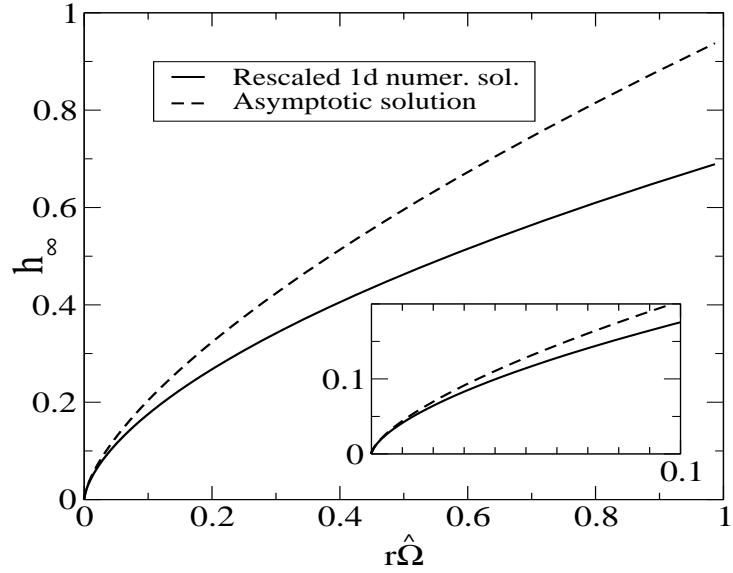}\qquad
\caption{Comparison of numerical results for the one-dimensional problem 
\rf{stateq} with the asymptotic formula for $h_{\infty}$ versus 
$r\hat\Omega$, \rf{hinfty2}.}
\label{fig:1dhinfty} 
\end{figure}

Figure \ref{fig:1dhinfty} shows $h_\infty$ as a function of 
$r\hat\Omega$. Good agreement of the one-dimensional numerical 
results with the corresponding higher order asymptotic formula 
is achieved for small values of $r\hat\Omega$.

\begin{figure}\centering
\includegraphics*[width=0.6\textwidth, height=0.5\textwidth]{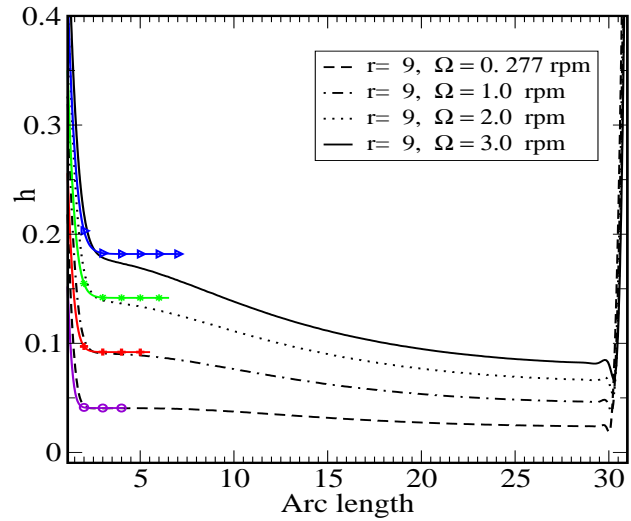}
\caption{Meniscus profiles  
computed with the 1D model \rf{stateq} (curves with symbols), 
for $\Omega=0.277\, ({\rm circles}), 1.0\, ({\rm stars}), 2.0\, ({\rm plusses}), 3.0\, ({\rm triangles})$, 
and comparison with the profiles obtained for the cross section 
of the film profile for the rotating disk. ``Arc length'' denotes $r\theta$.}
\label{fig:1dshoot} 
\end{figure}

The meniscus profile $h(y)$ in figure \ref{fig:1dshoot} is now computed for the 
values given in section (4.1.1).
Recall that $\hat{\Omega}=1.089e-3.$ At $r=9$, we have
$r \hat{\Omega}=0.009801$, hence, from \rf{hinfty2}, $h_\infty=0.0433$.

We now compare the meniscus profile computed with 
\rf{stateq2} for the drag-out problem with the numerical solution 
for the steady state 
of our problem \rf{profeq}--\rf{bcnat2}. For this 
we take results for the cross section along constant radii. 
In figure 6 we performed the comparison for the cross section 
for the height profile at radius 
$r=9$, for $0\leq\theta\leq 180$, i.e. from the point where the film is 
dragged out to the point where it reenters the liquid bath. 
We see that 
there is excellent quantitative agreement in the vicinity of the 
meniscus region as it enters the thin film region.

\subsubsection{Comparison with the hyperbolic regime}

Further out into the disk region the height profile will deviate 
from the height obtained for the drag-out problem.
There the variation of the height along the directions 
parallel to the disk is very small, 
which is clearly seen in our numerical simulations,  
so that surface tension will play a negligible role. 

Starting from equation \rf{profeq} we consider the steady state 
problem 
\be
\pr  \left[r\frac{h^3}{3} \sin\theta \right] 
+ \pth\left[\frac{h^3}{3} \cos\theta - \hat\Omega r h\right] = 0, 
\ee
to describe the dynamics far away from the meniscus.
This can be simplified to the hyperbolic equation 
\be
h^2 r \sin\theta\pd h r + \left(h^2\cos\theta-\hat\Omega r\right) \pd h\theta = 0
\ee

\begin{figure}\centering
\includegraphics*[width=0.7\textwidth]{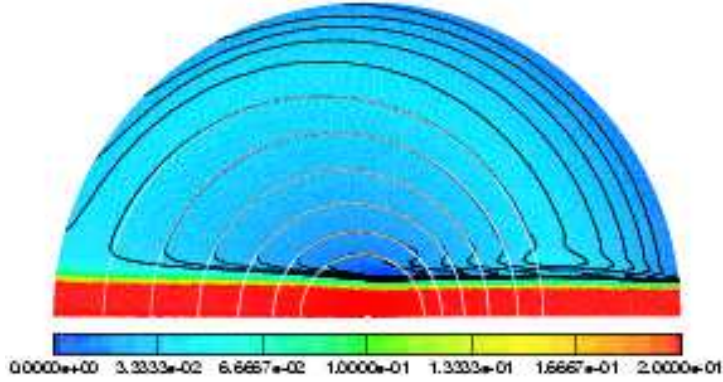}
\caption{Comparison of contour lines from our FEM computation (black curves) for $\Omega=0.277$ 
with the characteristics (white curves) for $r_0=1.962, 2.670, 3.560, 4.572, 5.759, 6.877$ and corresponding heights of 
$h_0=0.0130, 0.0182, 0.0234, 0.0286, 0.0338, 0.0390$, respectively. Note, the corresponding 3D plot in figure \ref{fig:HK1}.}
\label{fig:charac} 
\end{figure}

Using the method of characteristics, this problem can be solved in form 
of an intitial value problem for the system of the coupled 
ordinary differential equations 
\sbea
\frac{dr}{d\tau} &=& h^2(r_0,0)\, r\sin\theta, \qquad r(0)=r_0\label{ode1}\\
\frac{d\theta}{d\tau} &=& h^2(r_0,0)\cos\theta -\hat\Omega r, \qquad \theta(0)=0\label{ode2}
\seea
Using as the initial condition the height found from \rf{stateq2} or simply 
by making use of formula \rf{hinfty2} for a chosen $r_0$ we can integrate 
\rf{ode1}, \rf{ode2} to obtain characteristics. 
This is shown in figure \ref{fig:charac} 
for $\Omega=0.277$ as an example. The results are similar for the other 
angular velocities. As can be seen, the comparison of the 
characteristics that start from the meniscus region shows good agreement 
with the contour lines found from the FEM computation. 
Note, that the contour lines that start at the boundary of the 
rotating disk strongly depend on the conditions there. 

Interestingly, one can get a good idea on the film profile as a function of the 
angular velocity $\hat\Omega$ by simply solving \rf{ode1}, \rf{ode2} for 
$r$ as a function of $\theta$ directly by taking 
\be
\frac{dr/d\tau}{d\theta/d\tau} = \frac{dr}{d\theta} = 
\frac{r\sin\theta}{\cos\theta -\hat\Omega r/h_0^2},\label{ode3}
\ee
which can be solved to yield 
\be
r(\theta) = \frac{h_0^2}{\hat\Omega}\left(\cos\theta \pm \sqrt{\cos^2\theta - 2c_0\frac{\hat\Omega}{h_0^2}}\right),\label{ode3sol}
\ee
where the integration constant is 
\be
c_0 = r(0)\cos\theta(0)-\frac{\hat\Omega r^2(0)}{2 h^2_0}.\label{c0}
\ee

\subsection{Film patterns for slightly immersed disks}

When the disk is half-immersed (immersion height $a=0$) 
the film thickness  
above the drag-out region will be 
renewed with every rotation and can be described by the 
classic drag-out problem. Furthermore, the film profile is 
slightly deformed by the effects of gravity before being 
drawn into the liquid bath accompanied by a typical 
oscillation in the thicknes of the film. 
The simple pattern emerging from this configuration 
changes when the immersion height 
is $a$ is increased.

\begin{figure}\centering
\includegraphics*[width=0.47\textwidth]{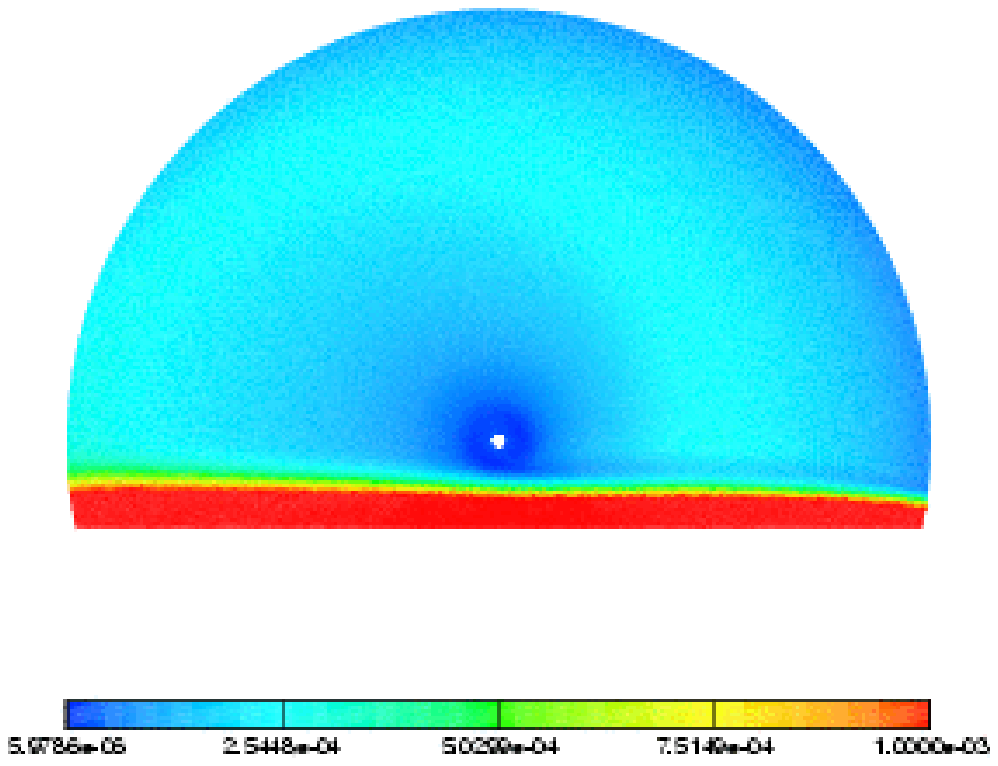}
\includegraphics*[width=0.47\textwidth]{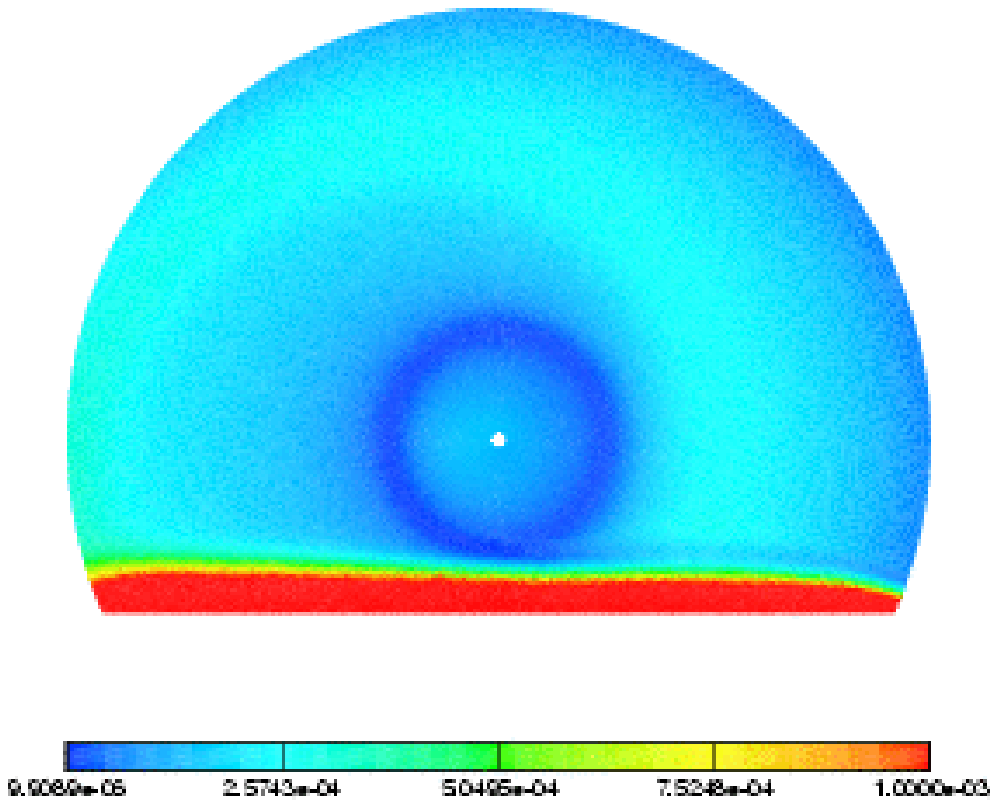}
\includegraphics*[width=0.47\textwidth]{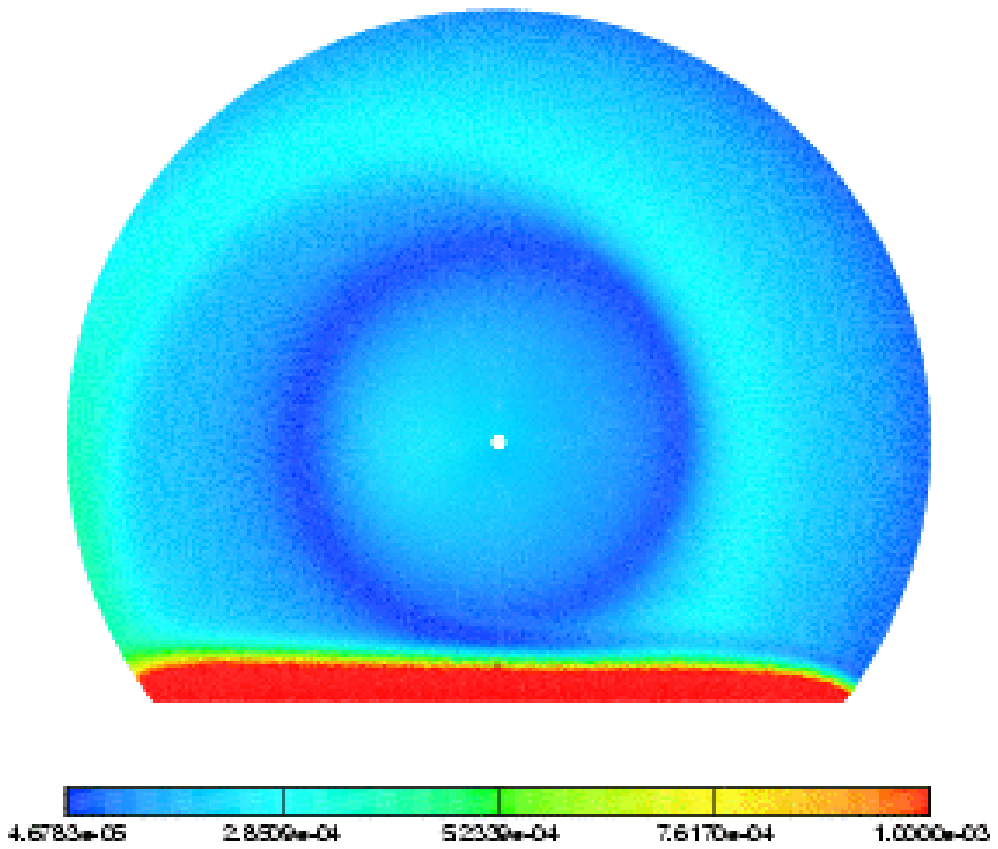}
\includegraphics*[width=0.47\textwidth]{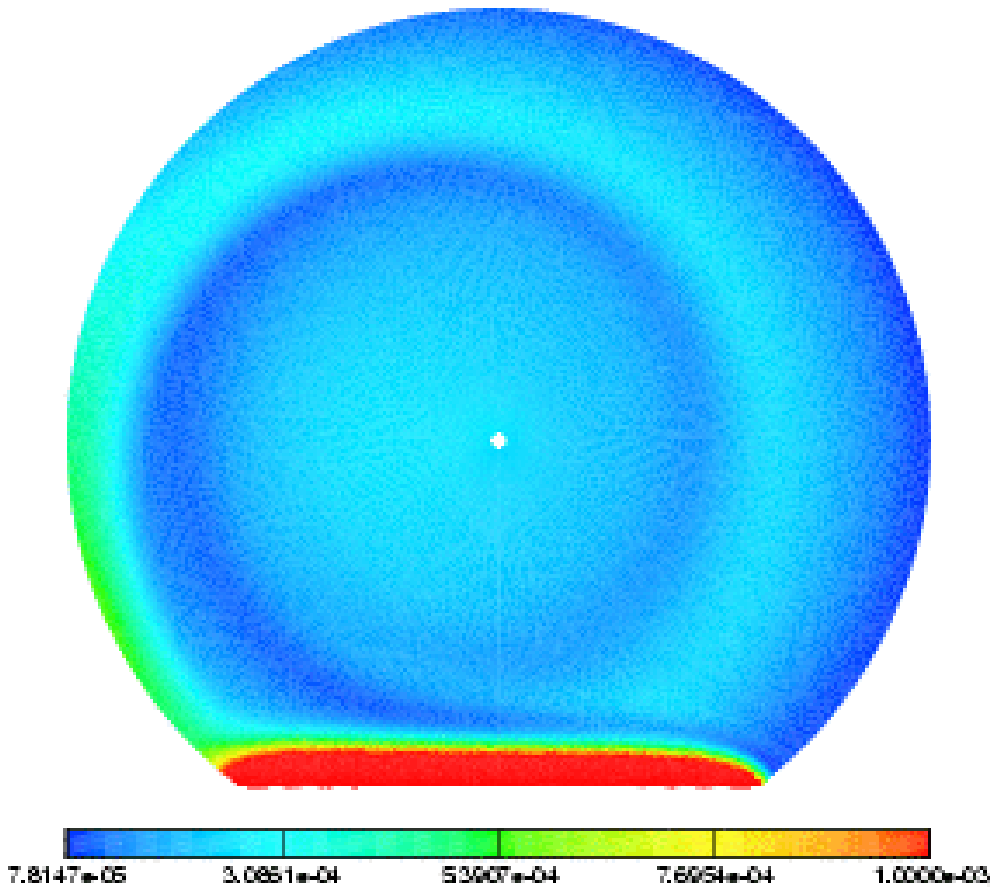}
\caption{Steady solution at $\Omega=1$\, r.p.m. for immersion depth 
of $a=0.2, 0.4, 0.6, 0.8$}
\label{fig:h1a0208}
\end{figure}

\begin{figure}\centering
\includegraphics*[width=0.4\textwidth,  height=0.3\textheight]{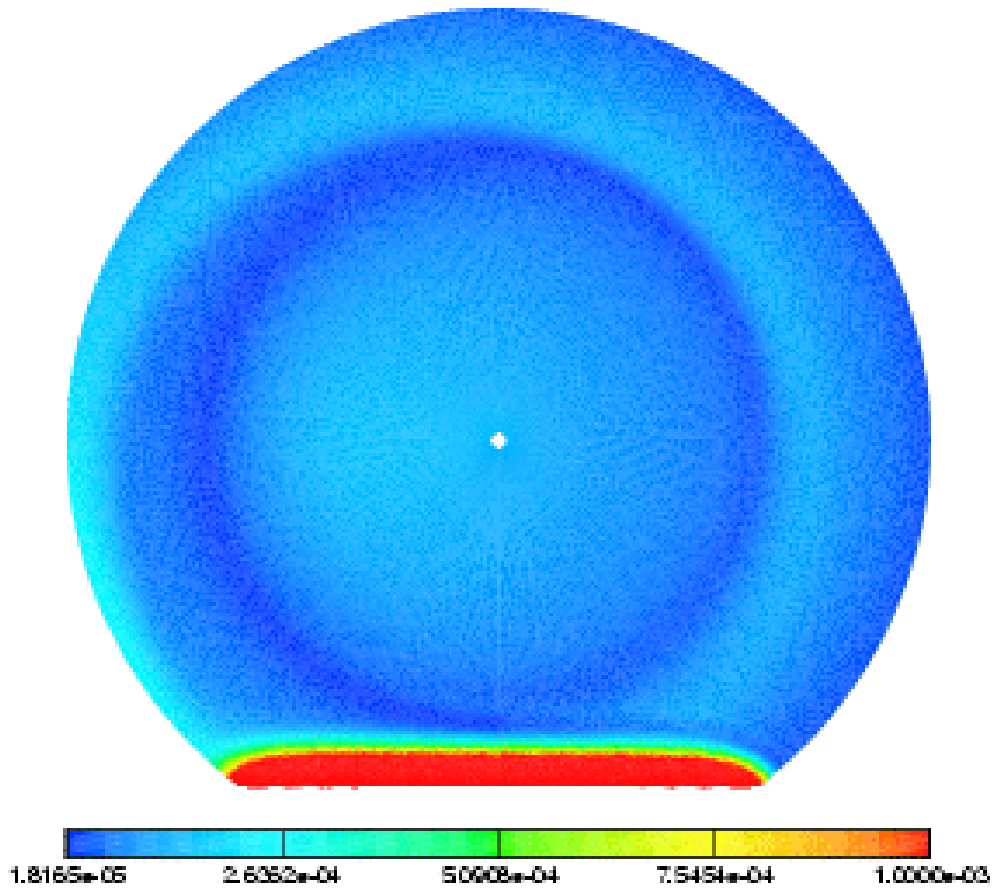}\\
\includegraphics*[width=0.4\textwidth, height=0.3\textheight]{h2D_oIL1.0_A0.8.eps}\\
\includegraphics*[width=0.4\textwidth, height=0.3\textheight]{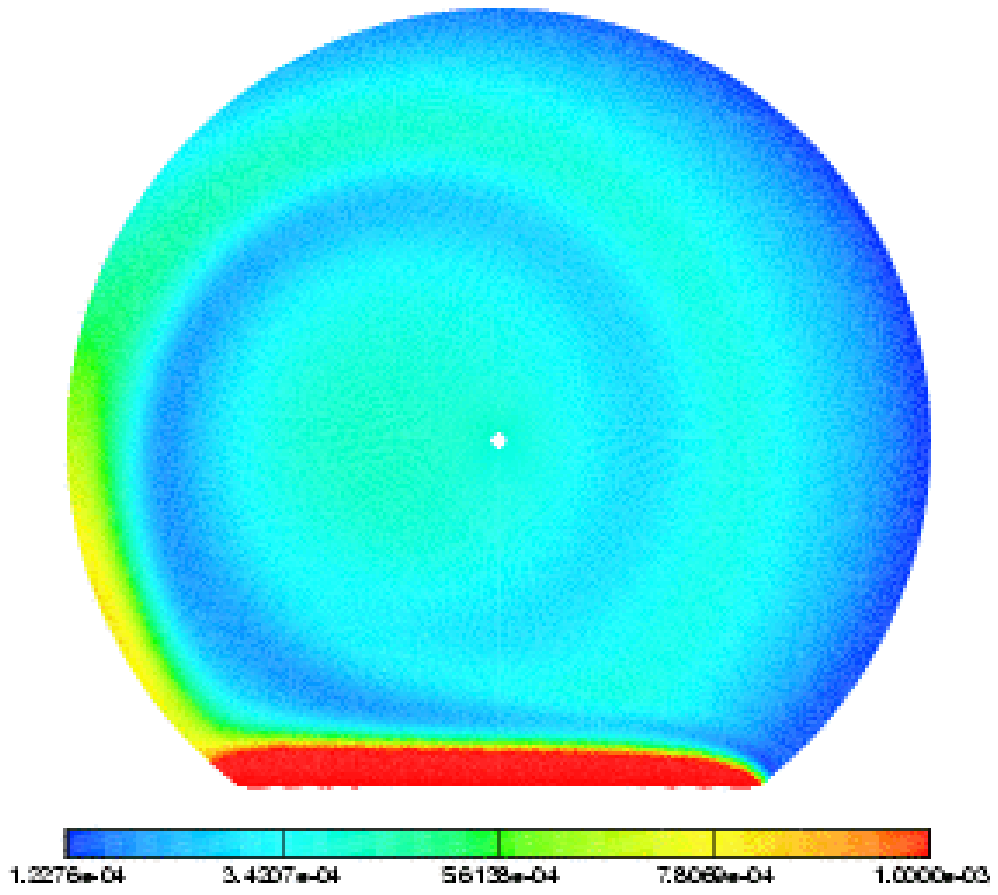}
\caption{Steady solution for immersion depth 
of $a=0.8$ at angular velocity $\Omega=0.277, 1.0, 2.0$\, r.p.m.}
\label{fig:h023a08}
\end{figure}

So in the final part of the paper we are interested in 
the new patterns that emerge when changing the two parameters 
$a$ and $\Omega$.

In the following figures \ref{fig:h1a0208} we 
first show the effect 
of varying the immersion depth from $a=0.2$ to $a=0.8$, leaving 
all other parameters as in the case of the half-immersed disk 
with $\Omega=1.0$. We observe an emerging almost symmetrical circular 
region where the film height has a minimum. This radius of the 
circular region increases with the immersion depth $a$. In fact, 
the radius quite closely corresponds to the distance from the minimum 
of the drag-in capillary wave to the vertical height of the 
rotation axis ($a=0$, see figure 1). By adjusting $a$ one 
can therfore achieve thin and fairly constant film profiles. 

\begin{figure}
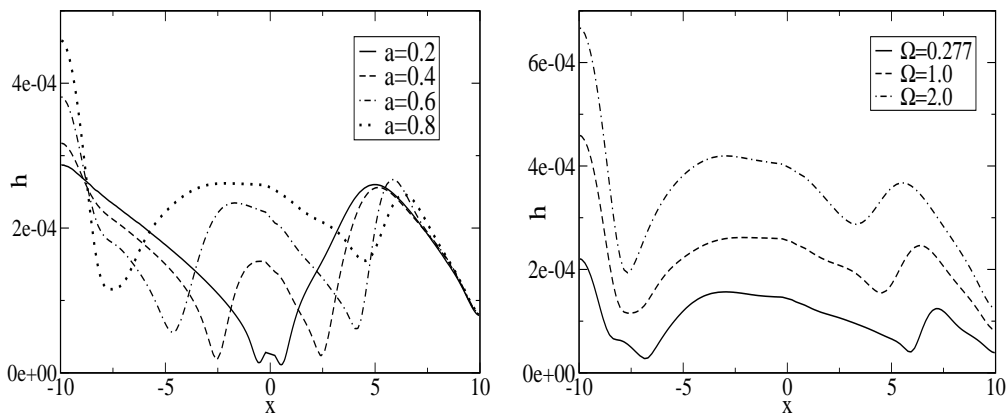
\centering
\includegraphics*[width=0.47\textwidth,height=0.28\textheight]{CrossSection_OM1.0.eps}\quad
\includegraphics*[width=0.47\textwidth,height=0.28\textheight]{CrossSection_A0.8.eps}\quad
\caption{Horizontal cross sections through $r=0$ for the film thickness 
as the immersion depth is increased (left) and as the angular velocity 
inincreased (right)}
\label{fig:fem-cross} 
\end{figure}

When the angular velocity is increased, not only does the fluid 
volume on the disk increase but the circular region is 
shifted towards the left, where the film is dragged out. This is demonstrated 
in figure \ref{fig:h023a08} for the case of the 
immersion height $a=0.8$. 
To see this effect more clearly we compare horizontal cross sections 
at $r=0$ for the film thickness, shown in figure \ref{fig:fem-cross}.

\section{Conclusions}
In this work we have derived a dimension-reduced generalized lubrication
model for the problem of the fully three-dimensional free-boundary problem
for the vertically rotating disk, dragging out a thin film from a liquid
bath.  The resulting two-dimensional nonlinear degenerate fourth-order
boundary value problem was solved numerically using a finite element
scheme. For the steady state solutions we performed an asymptotic analysis
near the meniscus region and a careful comparison with cross sections of
the numerical solutions along constant radii gave good agreement.
Away from the liquid bath we could show good agreement of our numerical
solution with analytic solutions of the corresponding hyperbolic problem.

For slightly immersed disks we observed patterns for the film
profile on the disk, which we studied as a function
of the immersion depth and the angular velocity. Interestingly, the
dependence of surface area of the film profile and of the volume fluid
dragged out is not trivial. 
Of course we only touched upon the rich structure of possible patterns 
and outlined some general tendencies. More detailed systematic study 
is subject of current research. It is now possible to find 
optimum configurations that will be
of importance for various technological and industrial applications. 
For example, in PET-reactors the fluid mechanical problem is coupled to
chemical reactions taking place mainly on the film surface, so that here
it is important to obtain thin films with high surface area.

\section*{Acknowledgements}

The authors acknowledge support by the DFG research
center {\sc Matheon}, Berlin and the DFG grant WA~1626/1-1.
AM also acknowledges support via the DFG grant MU~1626/3-1. 


\end{document}